\documentclass[10pt]{aastex62}

\usepackage{amsmath}
\usepackage{amssymb}
\usepackage{verbatim}

\graphicspath{{./}{figures/}}

\shorttitle{Tight bound on NS radius with QPOs in short GRBs}
\shortauthors{Guedes et al.}

\begin{document}

    \title{Tight bound on neutron-star radius with quasiperiodic oscillations in short gamma-ray bursts}
    
    \correspondingauthor{Victor Guedes}
    \author{Victor Guedes}
    \email{tpx5df@virginia.edu}
    \affiliation{Department of Physics, University of Virginia, Charlottesville, VA 22904, USA}

    \author{David Radice}
    \thanks{Alfred P.~Sloan Fellow}
    \affiliation{Institute for Gravitation \& the Cosmos, The Pennsylvania State University, University Park, PA 16802, USA}
    \affiliation{Department of Physics, The Pennsylvania State University, University Park, PA 16802, USA}
    \affiliation{Department of Astronomy \& Astrophysics, The Pennsylvania State University, University Park, PA 16802, USA}

    \author{Cecilia Chirenti}
    \affiliation{Department of Astronomy, University of Maryland, College Park, MD 20742, USA}
    \affiliation{Astroparticle Physics Laboratory, NASA/GSFC, Greenbelt, MD 20771, USA}
    \affiliation{Center for Research and Exploration in Space Science and Technology, NASA/GSFC, Greenbelt, MD 20771, USA}
    \affiliation{Center for Mathematics, Computation, and Cognition, UFABC, Santo André, SP 09210-170, Brazil}

    \author{Kent Yagi}
    \affiliation{Department of Physics, University of Virginia, Charlottesville, VA 22904, USA}

    \begin{abstract}
        Quasiperiodic oscillations (QPOs) have been recently discovered in the short gamma-ray bursts (GRBs) 910711 and 931101B. Their frequencies are consistent with those of the quasiradial and quadrupolar oscillations of binary neutron star merger remnants, as obtained in numerical relativity simulations.  These simulations reveal quasiuniversal relations between the remnant oscillation frequencies and the tidal coupling constant of the binaries.  Under the assumption that the observed QPOs are due to these postmerger oscillations, we use the frequency-tide relations in a Bayesian framework to infer the source redshift, as well as the chirp mass and the binary tidal deformability of the binary neutron star progenitors for GRBs 910711 and 931101B. We further use this inference to estimate bounds on the mass-radius relation for neutron stars. By combining the estimates from the two GRBs, we find a 68\% credible range $R_{1.4}=12.48^{+0.41}_{-0.40}$~km for the radius of a neutron star with mass $M=1.4$~M$_\odot$, which is one of the tightest bounds to date.
    \end{abstract}


    \section{Introduction}

        Recent measurements of neutron-star (NS) radii through X-ray observations~\citep{2019ApJ...887L..24M, 2021ApJ...918L..28M, 2019ApJ...887L..21R, 2021ApJ...918L..27R, 2024arXiv240614467D, 2024arXiv240614466S, 2024ApJ...971L..20C} have established significant constraints on the equation of state (EOS) for these objects, a long-standing problem in nuclear astrophysics. In addition, gravitational-wave (GW) observations from a binary neutron star (BNS) merger, GW170817~\citep{2017PhRvL.119p1101A}, and the concurrent detection of a short gamma-ray burst (GRB), GW170817A \citep{2017ApJ...848L..13A}, revolutionized multimessenger astronomy, constraining the EOS through measurements of the binary tidal deformability~\citep{Bauswein:2017vtn, Margalit:2017dij, 2018PhRvL.121p1101A, 2018PhRvL.121i1102D, 2018PhRvL.120z1103M, Radice:2017lry, Radice:2018ozg, 2020NatPh..16..907A, 2020NatAs...4..625C, 2020Sci...370.1450D, 2020ApJ...893L..21R, Breschi:2021tbm, 2022Natur.606..276H, Breschi:2024qlc} and strengthening the hypothesis that BNS mergers are sources of short GRBs.
        
        More recently, kilohertz quasiperiodic oscillations (QPOs) were found in the light curves of GRBs 910711 and 931101B \citep{2023Natur.613..253C}, raising the possibility that some short GRBs can exhibit sub-millisecond variability originating from the short GRB central engine. The central engines of short GRBs are required to launch a highly relativistic outflow, with postmerger BNSs being strong candidates, given the high energy and the rapid variability of the prompt emission~\citep{2007PhR...442..166N, 2014ARA&A..52...43B, 2023ApJ...958L..33G}.  Generally, the outcome of a BNS merger depends on the binary mass and on the EOS~\citep[see][for a recent review]{2021GReGr..53...59S}. 

        For example, low-mass BNSs can produce a stable NS remnant, while higher mass binaries can result in the prompt collapse to a black hole. Most relevant to our work, hypermassive neutron stars (HMNSs) are possible BNS merger remnants that are briefly supported against collapse to a black hole by differential rotation~\citep{2000ApJ...528L..29B}, lasting for only tens to hundreds of milliseconds, depending on how the stability of these objects is affected by thermal and magnetic processes after the merger~\citep{2020ARNPS..70...95R}. 

        Although a black hole central engine has been favored in the literature, the nature of the central engine and the necessary conditions for the successful launching and breakthrough of the jet are still under debate~\citep{2015JHEAp...7...73D, 2018IJMPD..2742004C}. For instance,~\cite{2014ApJ...788L...8M} suggests that the merger remnant must collapse within $\sim$ 100 ms, otherwise the jet is choked by the neutrino-driven wind from the remnant, thus supporting the prompt collapse and short-lived ($\lesssim 20$ ms) remnant scenarios. Nevertheless, observations of X-ray afterglows in short GRBs provide evidence that long-lived ($\gtrsim 20$ ms) remnants can produce the prompt emission and therefore power the afterglow~\citep{2015ApJ...805...89L} and this scenario is also supported by recent numerical relativity (NR) simulations~\citep{2020ApJ...901L..37M}. 
 
        The QPOs reported by~\cite{2023Natur.613..253C} have frequencies that can be interpreted as oscillation modes of the HMNS. GRBs 910711 and 931101B show two QPOs each, with similar centroid frequencies for both GRBs: a lower frequency QPO at $\sim 1$ kHz and a higher frequency QPO at $\sim 2.6$ kHz (see Table \ref{tab1}). NR simulations have shown that these BNS merger remnants have a rich GW spectrum, with peaks in the range $\sim$ 1 $-$ 5 kHz~\citep{2012PhRvL.108a1101B, 2015PhRvD..91f4001T}. The main peaks in the GW spectrum are characterized by the frequencies $f_{2}$ and $f_{2} \pm f_{0}$~\citep{2016PhRvD..93l4051R}, where $f_2$ is the frequency of the quadrupolar ($\ell=2$) mode, in the range $\sim$ 1 $-$ 4 kHz, and $f_{0}$ is the frequency of the quasiradial ($\ell=0$) mode\footnote{The ``$f_{0}$'' frequency should not be confused with an ``initial'' frequency or other spectral features of the postmerger GW signal, {\it e.g.}, the frequency at quasi time-symmetry in \cite{2024ApJ...960...86T}.}, around $\sim$ 1 kHz. We note that the frequencies of the QPOs reported in~\cite{2023Natur.613..253C} are in good agreement with these quadrupolar and quasiradial oscillation modes, thus supporting such an interpretation.

        If this correspondence holds, it is possible that the jet would have to be launched shortly after the merger, otherwise the modes would have been damped away due to the emission of GWs~\citep{2016PhRvD..94b4023B} and the modulation of the GRB signal would not be statistically significant~\citep{2019ApJ...884L..16C}. This suggests that the prompt emission should happen in the HMNS phase \citep[see][]{2024ApJ...961L..26C}. 

        The oscillation modes of short- or long-lived postmerger remnants have been shown to follow quasiuniversal relations ({\it i.e.}, relations that exhibit a weak dependence on the EOS) with premerger parameters such as the binary tidal deformability~\citep{2019PhRvD.100j4029B, 2021PhRvD.104d3011L, 2023CQGra..40h5011G}. These modes should be observed in the postmerger GW signal and their detection is expected to place tight bounds on NS radii and strong constraints on the EOS~\citep{2024arXiv240711153C}. Signatures of BNS postmerger remnants are still to be observed in GWs due to the lack of sensitivity of current detectors in the kilohertz range~\citep{2018LRR....21....3A}, a scenario that is expected to change with third generation detectors~\citep{2010CQGra..27s4002P, 2019BAAS...51g..35R}. Short GRBs, on the other hand, are detected $\sim$ weekly and the possibility that at least a fraction of these signals is modulated by the GW frequencies of the central engine brings a novel way to obtain information about the EOS of NS matter.

        In this work, we assume that the frequencies of the two QPOs observed in both GRB 910711 and GRB 931101B correspond to the frequencies of the quadrupolar mode $f_{2}$ and quasiradial mode $f_{0}$ from BNS simulations. Given this assumption, we obtain constraints on the redshift of the GRBs, as well as the binary tidal deformability and the chirp mass of the corresponding BNS merger event; we further estimate the radius of a 1.4 M$_{\odot}$ NS, obtain a constraint on the mass-radius diagram, and infer the masses and radii of the individual NSs before the merger.

    \section{Methods and Results}

        \label{mr}
    
        \subsection{Observations}
            
            We analyze GRBs 910711 and 931101B  separately. For each one, there are two measurements for the QPOs: $\nu_1 \pm \sigma_{\nu_1}$ and $\nu_2 \pm \sigma_{\nu_2}$, both measured in the detector frame and  thus redshifted relative to the source frame. We take the values for $\nu_1$ and $\nu_2$ from Table 1 of~\cite{2023Natur.613..253C}, and choose for $\sigma_{\nu_1}$ and $\sigma_{\nu_2}$ the average of the uncertainties for each measurement (a reasonable approximation since the reported $\pm1\sigma$ ranges are very similar). 

            In our analysis, we identify these QPOs observed in short GRBs with oscillation modes of HMNSs obtained from NR simulations of merging NSs. In short, we associate $\nu_2$ with the frequency of the quadrupolar ($\ell=2$) mode $f_2$ and $\nu_1$ with the quasiradial ($\ell=0$) mode $f_0$, both measured in the source frame (see Section~\ref{sim} for more details).  In order to keep a consistent notation (with respect to the modes that we are associating $\nu_{1}$ and $\nu_{2}$ with), we rename $\nu_1$ as $\nu_1 \equiv \nu_0$. Furthermore, for our purposes, it is more useful to work with $\nu_2$ and $\nu_2/\nu_0\equiv \nu_{02}$, which are our observables. Table~\ref{tab1} shows the measurements that we consider.

            \begin{table*}[t]
                \centering
                \begin{ruledtabular}
                    \begin{tabular}{cccccccc}
                         GRB & $T_{90}$ [ms] & $\nu_{0}$ [kHz] & $\sigma_{\nu_{0}}$ [kHz] & $\nu_{2}$ [kHz] & $\sigma_{\nu_{2}}$ [kHz] & $\nu_{02}$ & $\sigma_{\nu_{02}}$ \\
                         \hline
                         910711 & 14 & 1.113 & 0.008 & 2.649 & 0.007 & 2.38 & 0.02 \\
                         931101B & 34 & 0.877 & 0.007 & 2.612 & 0.009 & 2.98 & 0.03 
                    \end{tabular}
                    \caption{The QPO frequencies and associated uncertainties $\nu_{0}\pm\sigma_{\nu_{0}}$ and $\nu_{2}\pm\sigma_{\nu_{2}}$ for the GRBs 910711 and 931101B from~\cite{2023Natur.613..253C} (note that those authors use $\nu_{1}\pm\sigma_{\nu_{1}}$ to refer to the lower frequency QPO that, here, is $\nu_{0}\pm\sigma_{\nu_{0}}$). We consider $\nu_{2}$ and the ratio $\nu_{02}\equiv\nu_{2}/\nu_{0}$ as our measurements. We also quote $T_{90}$ for each GRB, which is the time span that contains 90$\%$ of the counts or, roughly, the duration of the GRB ({\it cf.} survival time of the remnant in Fig.~\ref{fig1}).}
                    \label{tab1}
                \end{ruledtabular}
            \end{table*}
    
    \subsection{Simulations}
    
        \label{sim}
    
        For given EOS and mass ratio $q \equiv M_1/M_2$ (for $M_1 \geq M_2$), we can label a BNS system by the chirp mass:
        \begin{align}
            \mathcal{M}\equiv\frac{(M_{1}M_{2})^{3/5}}{(M_{1}+M_{2})^{1/5}},
        \end{align}
        or the binary tidal deformability (see, {\it e.g.},~\citealt{2015PhRvD..91d3002L}):
        \begin{align}
            \tilde{\Lambda}\equiv\frac{16}{13}\left(\frac{(M_{1}+12M_{2})M^{4}_{1}{\bar{\Lambda}}_{1}}{(M_{1}+M_{2})^{5}}+(1\leftrightarrow2)\right),
        \end{align}
        where $M_A$ and ${\bar{\Lambda}}_A\equiv\Lambda_{A}/M^{5}_{A}$ are the mass and dimensionless tidal deformability of the $A^{\rm th}$ NS in the binary. The postmerger phase of BNS merger simulations is characterized by the amplitude of the GW signal $h(t)$ and the central rest-mass density of the postmerger remnant $\rho_{\rm max}(t)$, the latter being relevant in the case of a delayed collapse. The two dominant modes in the postmerger GW spectrum are the $\ell=2$ and $\ell=0$ modes, characterizing the frequencies $f_{2}$ and $f_{0}$. The latter is seen in the GW spectrum in the combination $f_2 \pm f_0$, but can also be extracted directly from the pulsations in the central density of the remnant\footnote{Indeed, this is the most reliable way to compute $f_{0}$ since the spectrum of the gravitational waveform from the simulations is sometimes ``noisy'' and have other peaks that can be misleadingly identified as $f_{2} \pm f_{0}$.}. Being the lowest order modes, these are the most likely to be imprinted in the electromagnetic counterpart of the GW signal and thus modulate the GRB signal.

        We consider a subset of the simulations of the CoRe (Computational Relativity) collaboration~\citep{2019PhRvD.100j4029B} that have been performed with the \texttt{WhiskyTHC} code and that report $\rho_{\rm max}$ for short- and long-lived remnants. We use these simulations to obtain the frequencies $f_{2}$ and $f_{0}$ by performing a Fourier analysis of $h(t)$ and $\rho_{\rm max}(t)$. More specifically, we use a set of 99 BNS simulations: 67 that result in long-lived remnants ({\it i.e.}, collapse does not happen within the time of the simulation) and 32 that result in short-lived remnants (collapse happens within the time of the simulation). There are also 53 simulations that result in prompt-collapse to a black hole and 12 simulations that have very few oscillation cycles in the central density and do not give a good estimate of $f_{0}$, neither of which we use in our analysis.

\begin{figure*}[t]
            \centering
            \includegraphics[width=\textwidth]{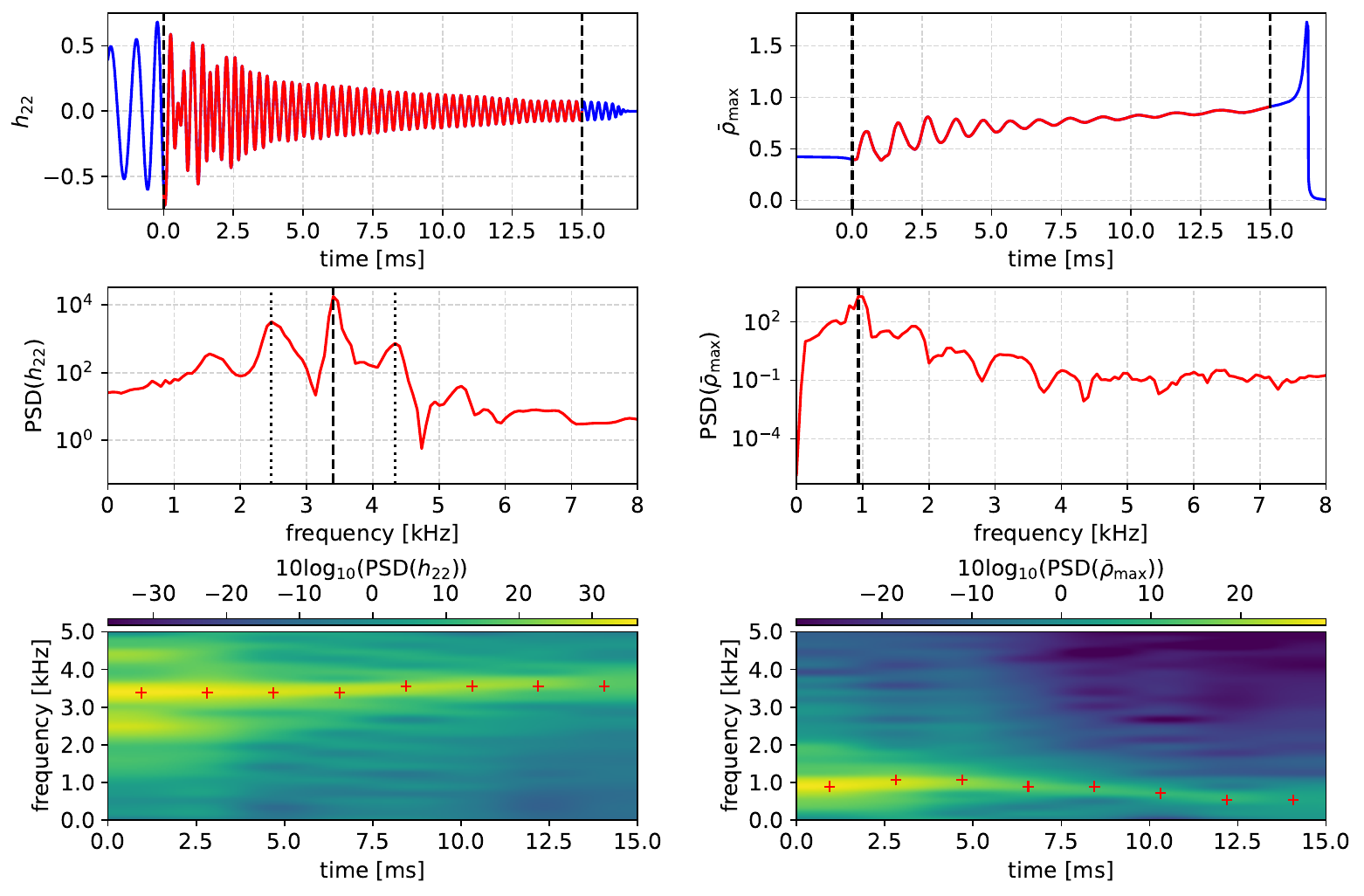}
            \caption{{\it Top}: $\ell=m=2$ waveform mode ($h_{22}$) for the postmerger GW signal (left) and the maximum rest-mass density for the postmerger remnant normalized by the maximum stable central density for a nonrotating star with that EOS (${\bar{\rho}}_{\rm max}\equiv\rho_{\rm max}/\rho_{\rm TOV}$) (right) from the merger of two NSs of mass $1.35$ M$_{\odot}$ described by the piecewise polytropic approximation to the SLy EOS~\citep{2009PhRvD..79l4032R}. The vertical dashed lines indicate the merger time $t_{\rm merger}$ and final time $t_{\rm final}$ within which we perform the Fourier transform. {\it Middle}: Power spectral density (PSD) of $h_{22}$ (left) and ${\bar{\rho}}_{\rm max}$ (right) between $t_{\rm merger}$ and $t_{\rm final}$. The vertical dashed lines indicate the peak frequencies in the spectra, $f_{2}$ (left) and $f_{0}$ (right), while the dotted lines in the left panel shows the beat frequencies $f_{2}\pm f_{0}$. {\it Bottom}: Spectrogram for $h_{22}$ (left) and ${\bar{\rho}}_{\rm max}$ (right). Note that the frequencies are approximately constant during the time span of the analysis and $f_{0}$ $\rightarrow$ $\sim 0$ as $t$ $\rightarrow$ $\sim t_{\rm final}$ as expected since the remnant collapses to a black hole.} 
            \label{fig1}
        \end{figure*}

        We compute the frequencies $f_{2}$ and $f_{0}$, from $h(t)$ and $\rho_{\rm max}(t)$, in the following way. For $h(t)$, we apply the Fourier transform between the time of merger, $t_{\rm merger}$ (for definitions, see~\citealt{2019PhRvD.100j4029B}), and the final time of the simulation, $t_{\rm final}$. In cases where the apparent horizon is formed within the time of the simulation, we redefine $t_{\rm final}$ for our analysis as the time shortly before when $\rho_{\max}(t)$ drops to zero (this indicates the collapse to a black hole as the region inside the apparent horizon is excluded from the analysis of the data). For the analysis of $\rho_{\rm max}(t)$, we detrend the function between $t_{\rm merger}$ and $t_{\rm final}$ before applying the Fourier transform by finding the best fourth order polynomial fit for $\rho_{\rm max}(t)$ and subtracting it from the function; we further remove the offset of the resulting function with respect to zero. We show in Fig.~\ref{fig1} an example of the Fourier transform for $h(t)$ and $\rho_{\rm max}(t)$, as well as their spectrograms, for a BNS system with $q=1$ ($M_{1}=M_{2}=1.35$ M$_{\odot}$) and the piecewise polytropic SLy EOS.

\begin{figure*}[t]
            \centering
            \includegraphics[width=\textwidth]{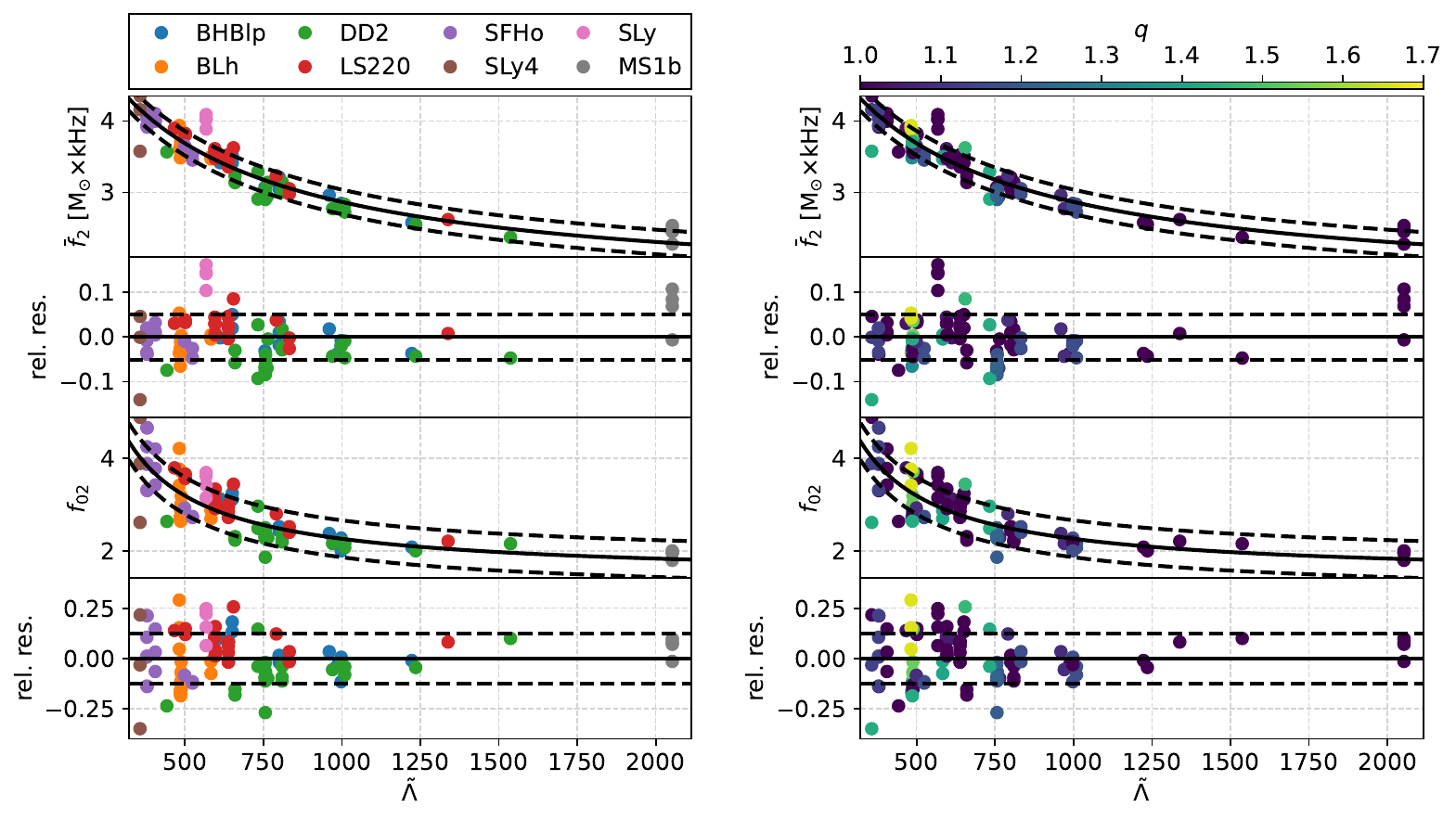}
            \caption{Quasiuniversal relations ${\bar{f}}_{2}(\tilde{\Lambda})$ and $f_{02}(\tilde{\Lambda})$. We show the EOS variation (left, see the different colors) and mass-ratio ($q$) variation (right, see the colorbar) of these relations. The solid lines are the best fits for these relations (using the fitting function in Eq.~\eqref{Qfit}) and the dashed lines represent the $1\sigma$ ($68.3\%$) credible regions. We also show the relative residuals $(Q_{i}-Q^{\rm fit}_{i})/Q^{\rm fit}_{i}$, where $Q$ $\in$ $\{$${\bar{f}}_{2}$, $f_{02}$$\}$, for these relations and the corresponding 1$\sigma$ credible regions.}
            \label{fig2}
        \end{figure*}

        Using the results for $f_{2}$ and $f_{0}$ for a range in $q$ between 1 and $\sim$ 1.66 and given the parameters $\mathcal{M}$ and $\tilde{\Lambda}$ for each binary, we obtain the new quasiuniversal relations (shown in Fig.~\ref{fig2}): $f_{2}/f_{0} \equiv f_{02}$ {\it vs.} $\tilde{\Lambda}$ and $\mathcal{M}f_{2} \equiv \bar{f}_{2}$ {\it vs.} $\tilde{\Lambda}$. Note that $f_{02}$ is independent of the redshift $z$, but $\bar{f}_{2}$ is not. We use a set of eight EOSs; six tabulated models: BHBlp~\citep{2014ApJS..214...22B}, BLh~\citep{2018A&A...609A.128B, 2010PhRvC..81a5803T}, DD2~\citep{2010PhRvC..81a5803T, 2010NuPhA.837..210H}, LS220~\citep{1991NuPhA.535..331L}, SFHo~\citep{2013ApJ...774...17S}, SLy4~\citep{2001A&A...380..151D}; and two piecewise polytropic fits to tabulated models: SLy and MS1b~\citep{2009PhRvD..79l4032R}. We construct fits for these quasiuniversal relations with the following fitting function \citep[inspired by Eq. (13) of][]{2019PhRvD.100j4029B} for a quantity~$Q$:
        \begin{align}
            Q^{\rm fit}(\tilde{\Lambda})=Q_{0}\left(\frac{1+n_{1}\tilde{\Lambda}+n_{2}{\tilde{\Lambda}}^{2}}{1+d_{1}\tilde{\Lambda}+d_{2}{\tilde{\Lambda}}^{2}}\right),
            \label{Qfit}
        \end{align}
        where $Q$ $\in$ $\{$${\bar{f}}_{2}$, $f_{02}$$\}$ while $Q_0$, $n_i$, and $d_i$ ($i$ $\in$ $\{$1, 2$\}$) are the fitting coefficients. We use least-squares regression to obtain the fitting coefficients for the relations $Q(\tilde{\Lambda})$ and compute their standard deviation, $\sigma_{Q}$, through:
        \begin{align}
            \sigma^{2}_{Q}=\frac{1}{N-1}\sum_{i=1}^{N}(Q_{i}-Q^{\rm fit}_{i})^{2}, \label{stdQ}
        \end{align}
        \noindent where $N$ is the number of data points ({\it i.e.}, 99 simulations, in our case). In Fig.~\ref{fig2}, we overlay the best fits for the relations along with the $\pm1\sigma$ credible regions and the residuals, where we show the EOS and mass-ratio variation of the relations separately. In Table~\ref{tab2}, we present the best-fit coefficients, the chi-square statistic, the Kullback-Leibler (KL) divergence between the distribution of the residuals and a normal distribution with zero mean and standard deviation $\sigma_{Q}$, and the average and maximum relative errors. The KL divergence is close to zero for both fits, which shows that the residuals are distributed according to a Gaussian distribution\footnote{We also used Bayesian regression to obtain alternative fits for the ${\bar{f}}_{2}(\tilde{\Lambda})$ and $f_{02}(\tilde{\Lambda})$ relations, assuming that the residuals are distributed according to a Gaussian and a Laplace distribution (which can be more appropriate when the data has outliers). We obtained that our least-squares fit is in good agreement with the Bayesian fit with a Gaussian distribution for the residuals, as we should expect. We also verified that no model for the distribution of the residuals is strongly preferred by the data. Thus, since the Gaussian distribution has the maximum entropy among other probability distributions and the corresponding Bayesian fit is in good agreement with the least-squares fit, we use the latter in our analysis.\label{note1}}.
        
        The quasiuniversal relations that we propose are slightly different from the ones reported in~\cite{2019PhRvD.100j4029B}. Those authors consider the relations ${\hat{f}}_{2}$ {\it vs.} $\xi$ and ${\hat{f}}_{2\pm0}$ {\it vs.} $\xi$. Their independent variable is a parameter $\xi$ that has dependence on the tidal polarizability $\kappa^{\rm T}_{2}$ and on the mass ratio. Their dependent variables are ${\hat{f}}_{2}$ and ${\hat{f}}_{2\pm0}$, which are the frequencies ${f}_{2}$ and ${f}_{2\pm0}$ normalized by the binary mass. The relations in this work do not depend explicitly on the mass ratio, which is not known for the hypothetical BNS progenitors of the GRBs under consideration. Furthermore, our relations are given in terms of the chirp mass and the binary tidal deformability (instead of the binary mass and the tidal polarizability), which are arguably the most relevant measurable mass and tidal parameters when inspiral GWs from BNSs are detected. We note that more recent fits between postmerger oscillation frequencies and premerger parameters have been proposed by, {\it e.g.}, \cite{2023CQGra..40h5011G, 2024PhRvD.109f4009B, 2024ApJ...960...86T}. However, these relations also have explicit mass-ratio and spin dependence. We assume that these effects are indirectly taken into account by the error of our relations.
        
        Finally, we note that our relations could be affected by phase transitions in the EOS and such effects have not been implemented in the set of simulations that we consider. However, we stress that such effects could be degenerate with the spread of the relations if not strong enough~\citep{2024PhRvD.109j3008P}.

        \begin{table*}[t]
            \centering
            \begin{ruledtabular}
                \begin{tabular}{cccccccccccc}
                    $Q$ & $Q_{0}$ & $n_{1}$ & $n_{2}$ & $d_{1}$ & $d_{2}$ & $\sigma_{Q}$ & $\chi^{2}$ & $D_{\rm KL}$ & avg$(\Delta Q$) & max($\Delta Q$) \\
                    \hline
                    ${\bar{f}}_{2}$ & 1.409 & 946.7 & 7.286$\times10^{-1}$ & 178.3 & 6.452$\times10^{-1}$ & 1.72$\times10^{-1}$ & 2.899 & 7.942$\times10^{-3}$ & 3.785$\times10^{-2}$ & 1.635$\times10^{-1}$ \\
                    $f_{02}$ & 2.161 & 5163 & 10.34 & $-$828.7 & 15.65 & 3.991$\times10^{-1}$ & 15.61 & 4.659$\times10^{-3}$ & 1.047$\times10^{-1}$ & 5.382$\times10^{-1}$
                \end{tabular}
                \caption{Fitting coefficients for the quasiuniversal relations ${\bar{f}}_{2}(\tilde{\Lambda})$ and $f_{02}(\tilde{\Lambda})$ using the fitting function in Eq.~\eqref{Qfit}. We also quote: the standard deviation for the parameters, $\sigma_{Q}$ (see Eq.~\eqref{stdQ}); the chi-square statistic, $\chi^{2}\equiv\sum^{N}_{i=1}r_{i}^{2}$, where $r_{i}=Q_{i}-Q^{\rm fit}_{i}$ are the residuals and $N$ is the number of data points ({\it i.e.}, 99 simulations, in our case); and the Kullback-Leibler divergence between the distribution $\mathcal{P}$ for the residuals and a normal distribution $\mathcal{Q}$ with zero mean and standard deviation $\sigma_{Q}$, $D_{\rm KL}\equiv\sum^{N}_{i=1}{\mathcal{P}_{i}}{\rm ln}(\mathcal{P}_{i}/\mathcal{Q}_{i})$, where $\mathcal{P}_{i}=\mathcal{P}(r_{i})$ and  $\mathcal{Q}_{i}=\mathcal{Q}(r_{i})$; the average and maximum relative error $\Delta Q = |1-Q^{\rm fit}/Q|$.}
                \label{tab2}
            \end{ruledtabular}
        \end{table*}

    \subsection{Priors, likelihoods, and posteriors}

        \label{pe}

        We use the best fits for the quasiuniversal relations $f_{02}(\tilde{\Lambda})$ and $\bar{f}_{2}(\tilde{\Lambda})$ and their standard deviations $\sigma_{f_{02}}$ and $\sigma_{\bar{f}_{2}}$ (see Table~\ref{tab2}), to obtain the joint posterior $P$ on $\tilde{\Lambda}$, $\mathcal{M}$, and $z$, by using the standard Bayesian expression (see, {\it e.g.},~\citealt{2020ApJ...888...12M} for application to EOS inference from NS observations):
        \begin{align}
            P(\tilde{\Lambda},\mathcal{M},z)\propto p(\tilde{\Lambda},\mathcal{M},z)\mathcal{L}(\tilde{\Lambda},\mathcal{M},z),
        \end{align}
       where $p({\tilde\Lambda},{\cal M},z)$ is the prior (normalized so that the integral over all values of ${\tilde\Lambda}$, ${\cal M}$, and $z$ is unity), ${\cal L}({\tilde\Lambda},{\cal M},z)$ is the likelihood of the data given the model, and the proportionality is because the posterior, like the prior, is a probability density and thus must be normalized.

        \paragraph{Prior} We decompose the prior as:
        \begin{align}
            p(\tilde{\Lambda},\mathcal{M},z)=p(\tilde{\Lambda},\mathcal{M})p(z),
        \end{align}
        \noindent where we take into account the correlation between $\tilde{\Lambda}$ and $\mathcal{M}$ (see, {\it e.g.},~\citealt{2018PhRvD..98f3020Z, 2022ApJ...939L..34A}) when writing the joint prior on these parameters, {\it i.e.},
        \begin{align}
            p(\tilde{\Lambda},\mathcal{M})=p(\tilde{\Lambda})p(\mathcal{M}|\tilde{\Lambda}).
        \end{align}
        \noindent For the range in $\mathcal{M}$ that we consider here ([1.04, 1.31] M$_{\odot}$), which is determined by our set of NR simulations, the relation $\mathcal{M}$ $vs.$ $\tilde{\Lambda}$ is approximately linear (with a Pearson correlation coefficient of $\sim$ $-0.5$). We can thus find the best fit $\mathcal{M}(\tilde \Lambda)=a_{0}+a_{1}\tilde{\Lambda}$ (with $a_{0} = 1.332$ M$_{\odot}$ and $a_{1} = -2.216\times10^{-4}$ M$_{\odot}$) and corresponding standard deviation ($\sigma_{\mathcal{M}}$ = $7.068\times10^{-2}$ M$_{\odot}$). As shown in~\cite{2022ApJ...939L..34A}, the relation $\mathcal{M}(\tilde{\Lambda})$ is generally nonlinear (see their Fig. 4), depending on the range of the variables. Since we only have eight EOSs in our sample, the nonlinear behavior of the relation is not obvious and thus, for practical purposes, we consider that the relation is approximately linear\footnote{We use least-squares regression to find the best linear fit for the $\mathcal{M}(\tilde{\Lambda})$ relation, but we also use Bayesian regression to obtain alternative fits, considering a linear model and the nonlinear model in~\cite{2022ApJ...939L..34A}. As expected, the least-squares fit agrees with the Bayesian linear fit, which is also in good agreement with the Bayesian nonlinear fit ({\it i.e.}, there is no strong preference between the models).\label{note2}}. Then, we can write:
        \begin{align}
            p(\mathcal{M}|\tilde{\Lambda})\propto{\rm exp}\left(-\frac{(\mathcal{M}-\mathcal{M}(\tilde{\Lambda}))^{2}}{2\sigma^{2}_{\mathcal{M}}}\right).
        \end{align}
        We assume $p(\tilde{\Lambda})$ to be uniform in the range informed by the NR simulations ([357, 2053]). We consider $p(z)$ to correspond to the redshift distribution in~\cite{2005A&A...435..421G}, obtained from the luminosity function determined from the peak flux distribution of short GRBs in the BATSE data. We consider the distribution that takes into account the star formation rate and the merger time delay of BNSs for short GRBs, with a peak in $z\sim 0.5$ (see their Fig.~3), and that agrees with more recent data (see, {\it e.g.}, Table 1 and Fig. 4 of~\citealt{2014ARA&A..52...43B} and Table 2 of~\citealt{2022ApJ...940...56F}); we consider the same range for $z$ ([0, 5]).
        
        \paragraph{Likelihood} The likelihood can be written as:
        \begin{align}
            \mathcal{L}(\tilde{\Lambda},\mathcal{M},z)=\mathcal{L}_{\nu_{02}}(\tilde{\Lambda})\mathcal{L}_{\nu_{2}}(\tilde{\Lambda},\mathcal{M},z),
        \end{align}
        \noindent where:
        \begin{align}
            \mathcal{L}_{\nu_{02}}(\tilde{\Lambda}) &\equiv \mathcal{L}(\nu_{02}|f_{02}(\tilde{\Lambda})) \propto {\rm exp}\left(-\frac{(\nu_{02}-f_{02}(\tilde{\Lambda}))^2}{2(\sigma^2_{\nu_{02}} + \sigma^2_{f_{02}})}\right), \label{Lnu02}
        \end{align}
        \noindent and
        \begin{align}
            \mathcal{L}_{\nu_{2}}(\tilde{\Lambda},\mathcal{M},z) &\equiv \mathcal{L}(\nu_{2}|f^{\rm obs}_{2}(\tilde{\Lambda},\mathcal{M},z)) \propto {\rm exp}\left(-\frac{(\nu_{2}-f^{\rm obs}_{2}(\tilde{\Lambda},\mathcal{M},z))^2}{2(\sigma^2_{\nu_{2}} + \sigma^2_{f^{\rm obs}_{2}}(\mathcal{M},z))}\right), \label{Lnu2}
        \end{align}
        \noindent {with the definitions} $f^{\rm obs}_{2}(\tilde{\Lambda},\mathcal{M},z) \equiv \bar{f}_{2}(\tilde{\Lambda})/\mathcal{M}(1+z)$ and $\sigma_{f^{\rm obs}_{2}}(\mathcal{M},z)\equiv\sigma_{{\bar{f}}_{2}}/\mathcal{M}(1+z)$. In Eqs.~\eqref{Lnu02} and \eqref{Lnu2}, we sum the standard deviations (from our model and the data) in quadrature because we assume that the uncertainties are uncorrelated with each other.

\begin{figure*}[t]
            \centering
            \includegraphics[width=\textwidth]{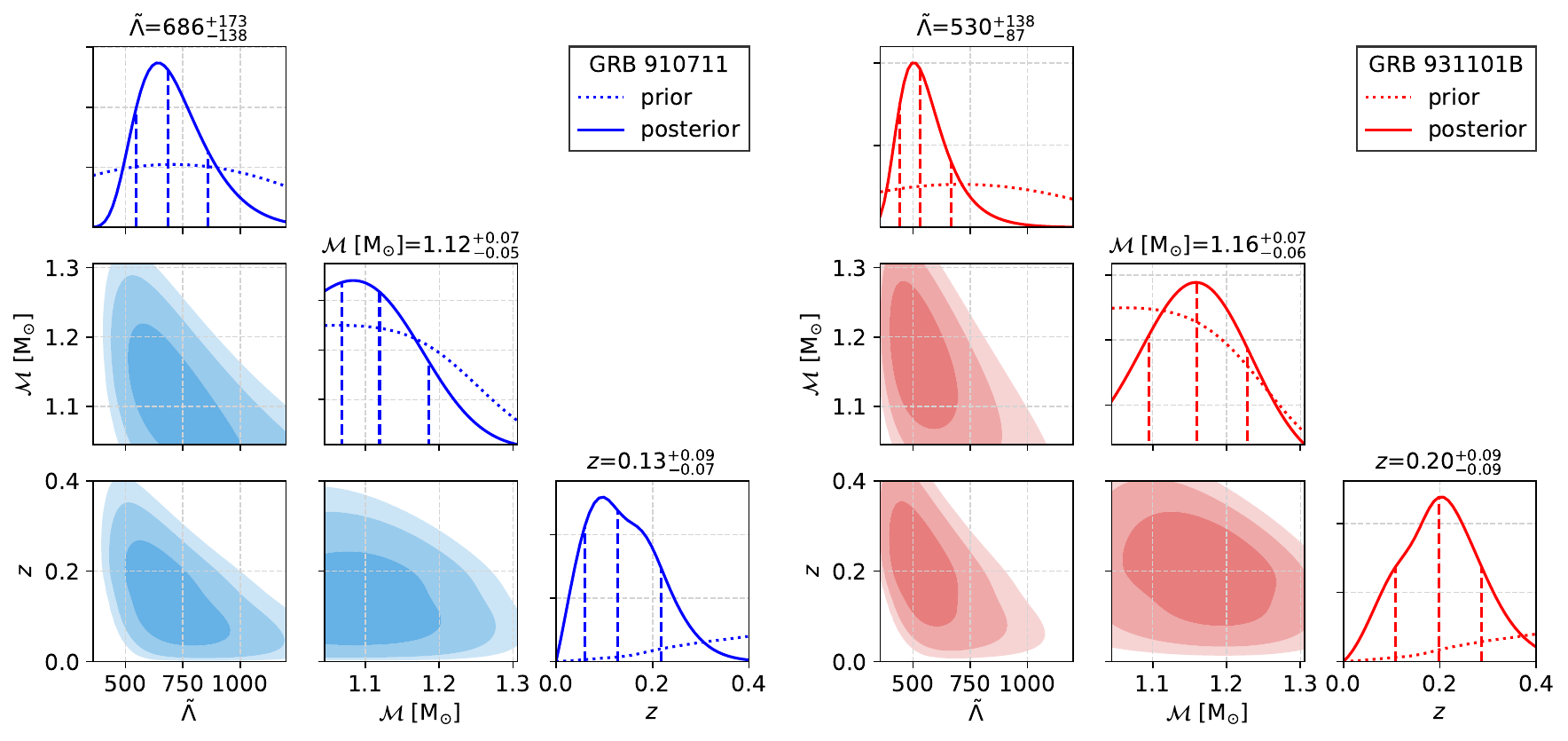}
            \caption{Results for the parameter estimation using the observed QPOs (see Table~\ref{tab1}) in GRBs 910711 (left) and 931101B (right), and the quasiuniversal relations ${\bar{f}}_{2}(\tilde{\Lambda})$ and $f_{02}(\tilde{\Lambda})$ (see Fig.~\ref{fig2}). {\it Off-Diagonal}: 2D marginalized posterior probability distributions for $(\tilde{\Lambda},\mathcal{M})$, $(\mathcal{M},z)$, and $(z,\tilde{\Lambda})$. The three different color tones represent the 1$\sigma$ (68.3$\%$), 2$\sigma$ (95.4$\%$), and 3$\sigma$ (99.7$\%$) credible regions from dark to light. {\it Diagonal}: 1D marginalized posterior probability distributions (solid) as well as prior (dotted) for $\tilde{\Lambda}$, $\mathcal{M}$, and $z$. The dashed lines represent the median and the $\pm1\sigma$ values.}
            \label{fig3}
        \end{figure*}

        In Fig.~\ref{fig3}, we show the results for the 1D and 2D marginalized posterior probability distributions for the parameters $\tilde{\Lambda}$, $\mathcal{M}$, and $z$ considering the measurements for the two GRBs in Table~\ref{tab1}. We summarize the results in the first three columns of Table~\ref{tab3}, where we quote the median and $\pm1\sigma$ ranges for the parameters. The posterior distributions are very informative, when compared to the prior distributions, with the values of $\tilde{\Lambda}$ and $\mathcal{M}$ being consistent with expected values for BNS mergers, and with the low values of $z$ being consistent with the expectation that GRBs 910711 and 931101B occurred relatively close to us~\citep{2023Natur.613..253C}.

        \begin{table*}[b]
            \centering
            \begin{ruledtabular}
                \caption{Results for the binary tidal deformability $\tilde{\Lambda}$, chirp mass $\mathcal{M}$, and redshift $z$ for the GRBs 910711 and 931101B. The values for $\tilde{\Lambda}$ and $\mathcal{M}$ are consistent with expected values for BNS mergers, {\it e.g.}, $\tilde{\Lambda}=390^{+500}_{-190}$ and $\mathcal{M}=1.118^{+0.001}_{-0.001}$ M$_{\odot}$ for GW170817~\citep{2019PhRvX...9a1001A}. We also quote the median and $\pm1\sigma$ ranges for the frequencies of the quadrupolar mode $f_{2}$, quasiradial mode $f_{0}$, and the ratio $f_{02}\equiv f_{2}/f_{0}$, that we determined through the result for $z$.}
                \begin{tabular}{ccccccc}
                    GRB & $\tilde{\Lambda}$ & $\mathcal{M}$ [M$_{\odot}$] & $z$ & $f_{2}$ [kHz] & $f_{0}$ [kHz] & $f_{02}$ \\
                    \hline
                    910711 & 686$^{+173}_{-138}$ & 1.12$^{+0.07}_{-0.05}$ & 0.13$^{+0.09}_{-0.07}$ & $3.00^{+0.22}_{-0.18}$ & $1.26^{+0.09}_{-0.07}$ & $2.39^{+0.23}_{-0.21}$ \\
                    931101B & 530$^{+138}_{-87}$ & 1.16$^{+0.07}_{-0.06}$ & 0.20$^{+0.09}_{-0.09}$ & $3.14^{+0.22}_{-0.22}$ & $1.05^{+0.07}_{-0.07}$ & $2.99^{+0.32}_{-0.29}$ \\
                \end{tabular}
                \label{tab3}
            \end{ruledtabular}
        \end{table*}

        In order to validate our results, we experimented with more conservative ranges for the priors in $\tilde{\Lambda}$ and $\mathcal{M}$ ([300, 5000] and [0.9, 1.4] M$_{\odot}$, respectively). The results for the posteriors are consistent with the previous results within the $\pm1\sigma$ ranges. We stress that the extended ranges for the priors rely on the assumption that the quasiuniversal relations, ${\bar{f}}_{2}(\tilde{\Lambda})$ and $f_{02}(\tilde{\Lambda})$ (see Fig.~\ref{fig1}), and the correlation between $\mathcal{M}$ and $\tilde{\Lambda}$ can be extrapolated. 
        
        We also tried using different fitting expressions for the relations $\mathcal{M}(\tilde{\Lambda})$, ${\bar{f}}_{2}(\tilde{\Lambda})$, and $f_{02}(\tilde{\Lambda})$ (see Footnotes~\ref{note1} and~\ref{note2} and corresponding discussion in the main text), and different redshift distributions for short GRBs as priors for $z$: a distribution that does not take into account the delay time for BNS mergers \citep[see Fig. 3 of][]{2005A&A...435..421G}, the data distribution in~\cite{2014ARA&A..52...43B}, and the recent data distribution in~\cite{2022ApJ...940...56F}. We found that the results are consistent with the results in Table~\ref{tab3} within the $\pm1\sigma$ ranges. 

        \subsection{Estimating the source-frame frequencies, constraining the neutron-star mass-radius relation, and inferring the individual masses and radii}

        Based on the inferences for $\tilde \Lambda$, $\mathcal M$, and $z$, we now determine the source-frame frequencies $f_{\ell}$ ($\ell$ $\in$ $\{$0, 2$\}$), which are:
        \begin{align}
            f_{\ell}=f^{\rm obs}_{\ell}(1+z).
            \label{sff}
        \end{align}
        From the parameter estimation performed in Section~\ref{pe},  we have $P(\tilde{\Lambda},\mathcal{M},z)$, and we can obtain the distribution for the redshift $z$ by marginalizing $P(\tilde{\Lambda},\mathcal{M},z)$ over $\tilde{\Lambda}$ and $\mathcal{M}$:
        \begin{align}
            P(z)=\int P(\tilde{\Lambda},\mathcal{M},z){\rm d}\tilde{\Lambda}{\rm d}\mathcal{M}.
        \end{align}
        \noindent From Eq.~\eqref{sff}, $f_{\ell}=f_{\ell}(f^{\rm obs}_{\ell},z)$, which we can invert:
        \begin{align}
            z=z(f_{\ell},f^{\rm obs}_{\ell})=\frac{f_{\ell}}{f^{\rm obs}_{\ell}}-1,\nonumber
        \end{align}
        then,
        \begin{align}
            &P(f_{\ell}){\rm d}f_{\ell}=P(z){\rm d}z \Rightarrow\nonumber\\ &P(f_{\ell}|f^{\rm obs}_{\ell})=P(z(f_{\ell},f^{\rm obs}_{\ell}))\frac{\partial{z}}{\partial{f_{\ell}}}=P(z(f_{\ell},f^{\rm obs}_{\ell}))\frac{1}{f^{\rm obs}_{\ell}},\nonumber
        \end{align}
        and thus,
        \begin{align}
            P(f_{\ell})=\int P(f_{\ell}|f^{\rm obs}_{\ell}){\rm d}f^{\rm obs}_{\ell}=\int P(z(f_{\ell},f^{\rm obs}_{\ell}))\frac{1}{f^{\rm obs}_{\ell}}P(f^{\rm obs}_{\ell}){\rm d}f^{\rm obs}_{\ell}, \label{Pfl}
        \end{align}
        \noindent where $P(f^{\rm obs}_{\ell})$ is the posterior for $f^{\rm obs}_{\ell}$, which is determined from the measurement of $\nu_{\ell}$ (see Table~\ref{tab1}), {\it i.e.}, it is a normal distribution with mean $\nu_{\ell}$ and standard deviation $\sigma_{\nu_{\ell}}$. We thus find:
        \begin{align}
            P(f_{\ell})=\int \frac{1}{\sqrt{2\pi}\sigma_{\nu_{\ell}}}{\rm exp}\left(-\frac{(f^{\rm obs}_{\ell}-\nu_{\ell})^{2}}{2\sigma^{2}_{\nu_{\ell}}}\right)P(z(f_{\ell},f^{\rm obs}_{\ell}))\frac{1}{f^{\rm obs}_{\ell}}{\rm d}f^{\rm obs}_{\ell}.
        \end{align}
        We show the results for the median and $\pm1\sigma$ ranges for the frequencies $f_{2}$ and $f_{0}$, and the ratio $f_{02}$, in the last three columns of Table~\ref{tab3}. These are the first measurements of the intrinsic oscillation frequencies of a short-lived BNS merger remnant, given our working assumption for the identification of the GRB QPOs.

        We next determine the radius $R_{M}$ of a NS of mass $M$ and constrain the mass-radius relation. We use the quasiuniversal relations proposed in~\cite{2021Univ....7..368G}: 
        \begin{align}
            R_{M}=\alpha\left(\frac{\mathcal{M}}{1\textrm{ }{\rm M}_{\odot}}\right)\left(\frac{\tilde{\Lambda}}{800}\right)^{\frac{1}{\beta}}, \label{RM}
        \end{align}
        where $\alpha$ (which is in km units) and $\beta$ are functions of $\mathcal{M}$ and $M$, and $M$ $\in$ $[$1.4, 2.14$]$ M$_{\odot}$. These relations are quasiuniversal with respect to the EOS and the mass ratio of the BNS, similar to the relations that we construct here (see Fig.~\ref{fig1}). We can obtain the joint posterior for $\tilde{\Lambda}$ and $\mathcal{M}$ by marginalizing $P(\tilde{\Lambda},\mathcal{M},z)$ over $z$:
        \begin{align}
            P(\tilde{\Lambda},\mathcal{M})=\int  P(\tilde{\Lambda},\mathcal{M},z){\rm d}z.
        \end{align}
        From Eq.~\eqref{RM}, $R_{M}=R_{M}(\tilde{\Lambda},\mathcal{M})$, which we can invert:
        \begin{align}
            \tilde{\Lambda}=\tilde{\Lambda}(R_{M},\mathcal{M})=800\left[\left(\frac{R_{M}}{\alpha}\right)\left(\frac{1\textrm{ }{\rm M}_{\odot}}{\mathcal{M}}\right)\right]^{\beta},\nonumber
        \end{align}
        thus,
        \begin{align}
            &P(R_{M},\mathcal{M}){\rm d}R_{M}{\rm d}\mathcal{M}=P(\tilde{\Lambda},\mathcal{M}){\rm d}\tilde{\Lambda}{\rm d}\mathcal{M} \Rightarrow\nonumber\\ &P(R_{M},\mathcal{M})=P(\tilde{\Lambda}(R_{M},\mathcal{M}),\mathcal{M})\frac{\partial{\tilde{\Lambda}}}{\partial{R_{M}}}=P(\tilde{\Lambda}(R_{M},\mathcal{M}),\mathcal{M})\frac{800\beta}{R_{M}}\tilde{\Lambda}(R_{M},\mathcal{M}),\nonumber
        \end{align}
        \noindent and then,
        \begin{align}
            P(R_{M})=\int P(\tilde{\Lambda}(\mathcal{M},R_{M}),\mathcal{M})\frac{800\beta}{R_{M}}\tilde{\Lambda}(R_{M},\mathcal{M}){\rm d}\mathcal{M}.
        \end{align}

        \begin{figure*}[t]
            \centering
            \includegraphics[width=\textwidth]{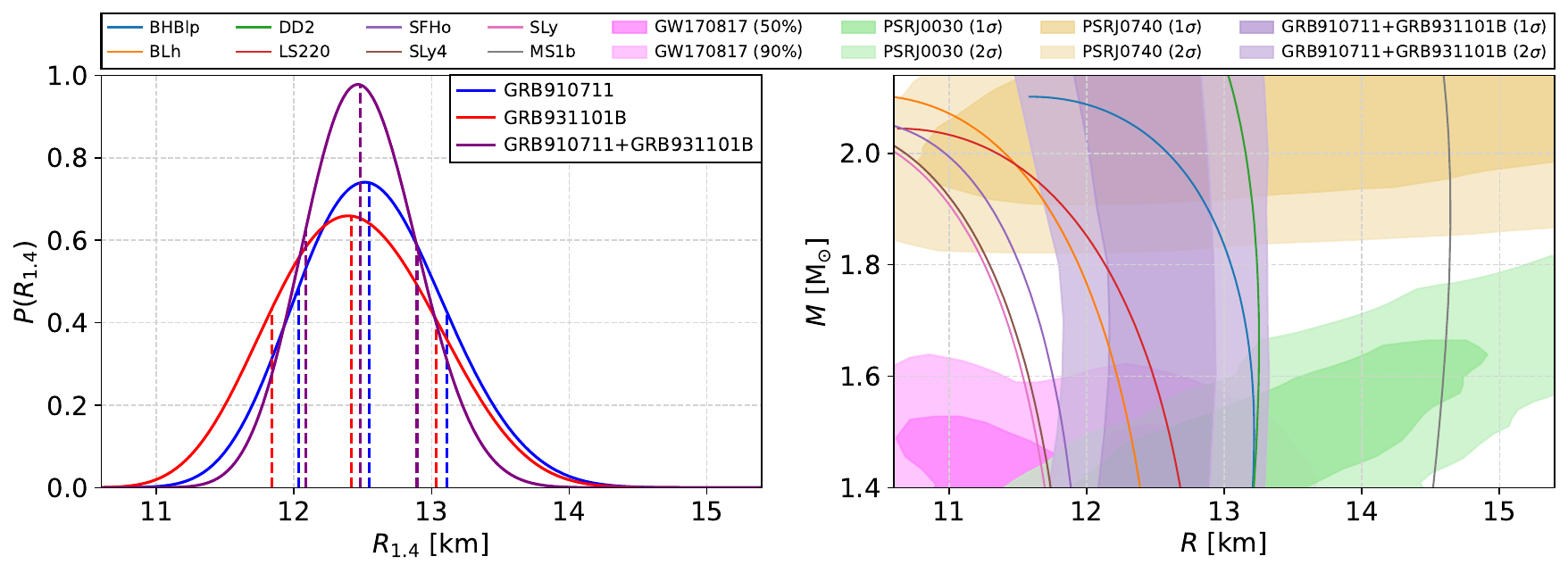}
             \caption{{\it Left}: Posterior probability distributions for the radius of a 1.4 M$_{\odot}$ NS for GRB 910711, GRB 931101B, and both. The dashed lines represent the median and the $\pm1\sigma$ values. {\it Right}: Credible regions (1$\sigma$ and 2$\sigma$) on the mass-radius plane from the constraint for both GRBs. We also show the mass-radius curves for the EOSs used in the NR simulations, and the credible regions for the mass-radius measurements obtained from GW observations for GW170817~\citep{2018PhRvL.121p1101A}, using parameterized EOS inference, and X-ray observations for PSR~J0030+0451 ({\it e.g.},~\citealt{2019ApJ...887L..24M}) and PSR~J0740+6620 ({\it e.g.},~\citealt{2021ApJ...918L..28M}).} 
            \label{fig4}
        \end{figure*}

        \begin{table*}[b]
            \centering
            \begin{ruledtabular}
                \caption{Inferred radii of NSs with mass in the range [1.4, 2.14] M$_{\odot}$ (restricted from the validity of the quasiuniversal relations in~\citealt{2021Univ....7..368G}), that we determined with the binary tidal deformability $\tilde{\Lambda}$ and the chirp mass $\mathcal{M}$ for each GRB and both. The value for $R_{1.4}$ is consistent with current bounds, {\it e.g.}, $R_{1.4}=12.57^{+0.49}_{-0.48}$ km~\citep{2024arXiv240614467D}.}
                \begin{tabular}{ccccccccc}
                    GRB & $R_{1.4}$ [km] & $R_{1.5}$ [km] & $R_{1.6}$ [km] & $R_{1.7}$ [km] & $R_{1.8}$ [km] & $R_{1.9}$ [km] & $R_{2.0}$ [km] & $R_{2.14}$ [km] \\
                    \hline
                    910711 & $12.55^{+0.56}_{-0.51}$ & $12.59^{+0.53}_{-0.49}$ & $12.60^{+0.54}_{-0.49}$ & $12.60^{+0.53}_{-0.48}$ & $12.58^{+0.54}_{-0.50}$ & $12.53^{+0.57}_{-0.53}$ & $12.49^{+0.60}_{-0.57}$ & $12.41^{+0.64}_{-0.61}$ \\
                    931101B & $12.42^{+0.62}_{-0.58}$ & $12.47^{+0.58}_{-0.57}$ &  $12.49^{+0.59}_{-0.54}$ & $12.50^{+0.57}_{-0.54}$ & $12.50^{+0.57}_{-0.56}$ & $12.45^{+0.60}_{-0.58}$ & $12.41^{+0.64}_{-0.61}$ & $12.33^{+0.68}_{-0.64}$ \\
                    910711+931101B & $12.48^{+0.41}_{-0.40}$ & $12.52^{+0.39}_{-0.39}$ & $12.54^{+0.40}_{-0.38}$ & $12.55^{+0.38}_{-0.38}$ & $12.53^{+0.39}_{-0.38}$ & $12.48^{+0.43}_{-0.40}$ & $12.44^{+0.44}_{-0.42}$ & $12.36^{+0.47}_{-0.45}$ \\
                \end{tabular}
                \label{tab4}
            \end{ruledtabular}
        \end{table*}

        \noindent This posterior for $R_{M}$ does not take into account the uncertainty in $R_{M}$ from the spread of the quasiuniversal relations in Eq.~\eqref{RM}. We account for this EOS and mass-ratio variation by assuming that each $R_{M}$ is the mean of a normal distribution with standard deviation $\sigma_{R_{M}}$, that is provided by~\cite{2021Univ....7..368G} as a function of $\mathcal{M}$ (see their Fig.~7). We thus have:
        \begin{align}
            P(R_{M})=\iint\frac{1}{\sqrt{2\pi}\sigma_{R_{M}}(\mathcal{M})}{\rm exp}\left(-\frac{(R^{\prime}_{M}-R_{M})^{2}}{2\sigma^{2}_{R_{M}}(\mathcal{M})}\right)P(\tilde{\Lambda}(R^{\prime}_{M},\mathcal{M}),\mathcal{M})\frac{800\beta}{R^{\prime}_{M}}\tilde{\Lambda}(R^{\prime}_{M},\mathcal{M})p(R^{\prime}_{M}){\rm d}R^{\prime}_{M}{\rm d}\mathcal{M},
        \end{align}
        where $p(R^{\prime}_{M})$ is the prior distribution for $R^{\prime}_{M}$, which we consider to be uniform. In the left panel of Fig.~\ref{fig4}, we show the estimate of the radius of a $1.4$ M$_{\odot}$ NS and the credible regions on the mass-radius plane, considering the measurements for the two GRBs in Table~\ref{tab1}. We quote the mean and $\pm1\sigma$ ranges for the radii (for $M$ $\in$ $[$1.4, 2.14$]$ M$_{\odot}$) in Table~\ref{tab4}.

        Considering the two GRBs, we infer $R_{1.4}=12.48^{+0.41}_{-0.40}$ km, which, under our assumptions, is one of the tightest estimates of $R_{1.4}$ to date, besides being consistent with current estimates, {\it e.g.},~\cite{2024arXiv240614467D} reports $R_{1.4}=12.57^{+0.49}_{-0.48}$ km (based on constraints from nuclear experiments, high-mass pulsars, LIGO data, NICER data, and EOS modelling using Gaussian processes). We report $\pm1\sigma$ ranges for our estimate, but our $\pm2\sigma$ range is $R_{1.4}=12.48^{+0.81}_{-0.76}$km, which is consistent with and as tight as recent multimessenger studies, {\it e.g.},~\cite{2024arXiv240204172K} reports $R_{1.4}=12.26^{+0.80}_{-0.91}$ km (based on constraints from nuclear theory, heavy pulsars, LIGO data, and NICER data). The mass-radius constraint, shown in the right panel of Fig.~\ref{fig4}, is consistent with mass-radius measurements for GW170817~\citep{2018PhRvL.121p1101A} as well as for PSR~J0030+0451 ({\it e.g.},~\citealt{2019ApJ...887L..24M}) and PSR~J0740+6620 ({\it e.g.},~\citealt{2021ApJ...918L..28M}).

        \begin{figure*}[t]
            \centering
            \includegraphics[width=0.45\textwidth]{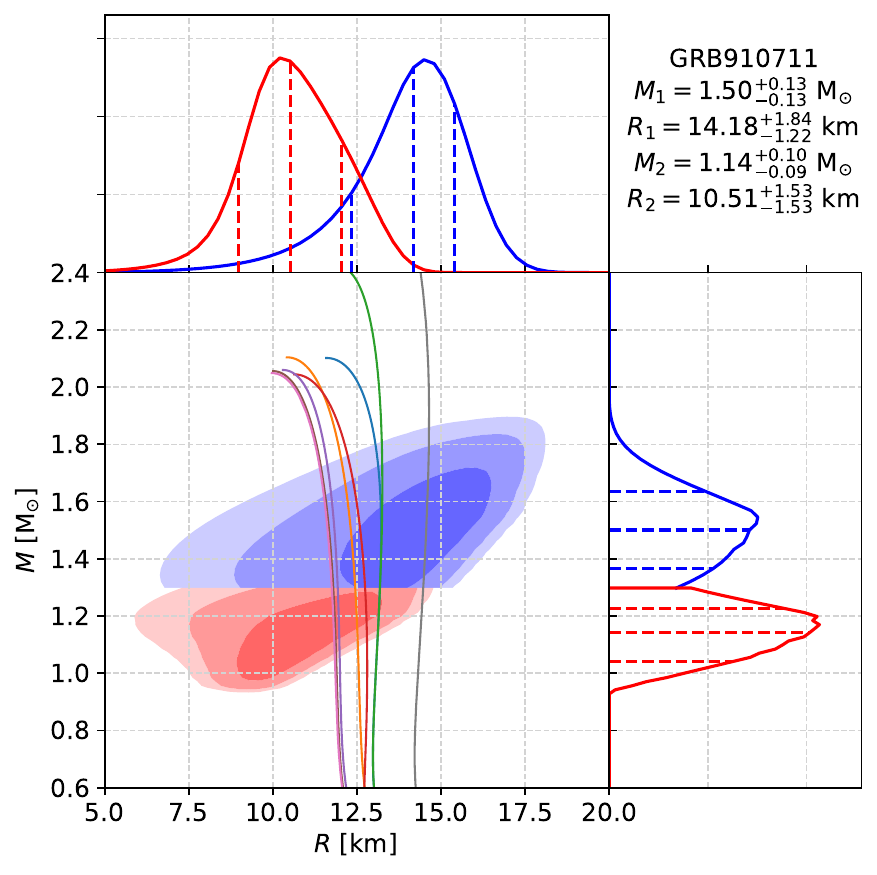}
            \includegraphics[width=0.45\textwidth]{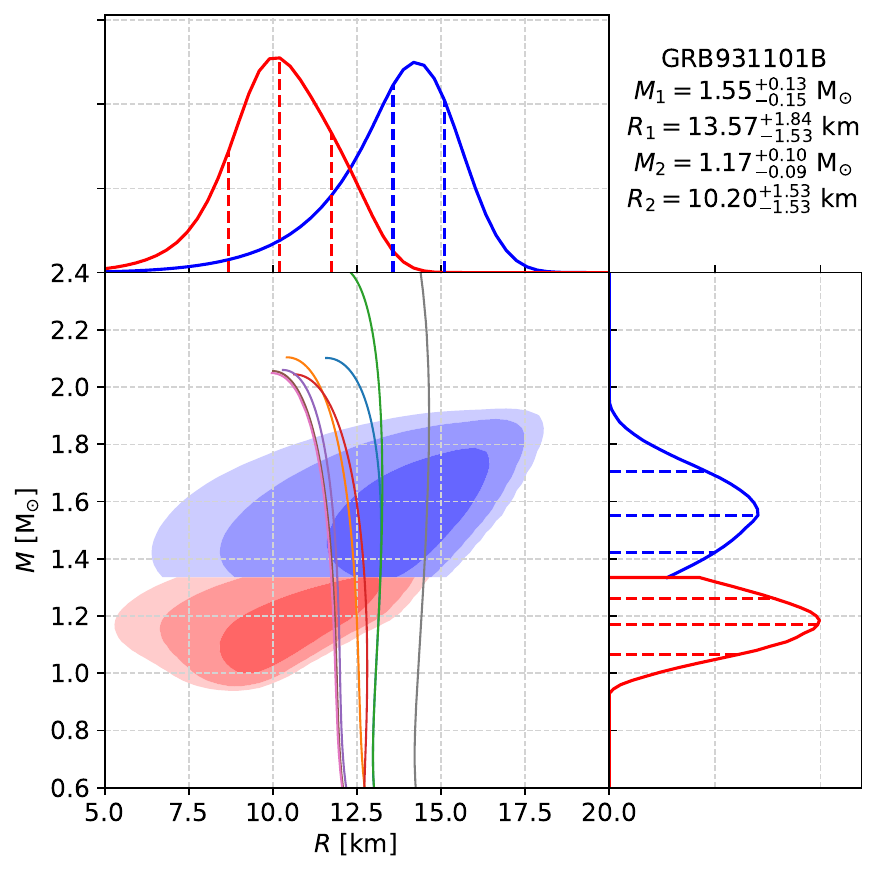}
             \caption{Mass-radius posterior distributions for the individual stars in the BNS systems which, according to our interpretation, produced the GRBs 910711 and 931101B. We show the 1$\sigma$, 2$\sigma$, and 3$\sigma$ credible regions (from dark to light) for the primary (blue) and secondary (red) stars. We also show the marginalized distributions for mass and radius in the top and right panels, where the dashed lines indicate the median and $\pm1\sigma$ values. We report the numerical results explicitly in the top right. The mass-radius curves for the EOSs are the same as in Fig.~\ref{fig4}.} 
            \label{fig5}
        \end{figure*}

        Our interpretation of GRBs 910711 and 931101B as being produced by BNS mergers allows us to infer the properties of the binary, such as their chirp mass and binary tidal deformability. With this information, we are able to determine the properties of the individual stars before the merger, given their mass ratio and considering the quasiuniversal Love-C relation in~\cite{2018PhRvD..98f3020Z} (see also~\citealt{2013PhRvD..88b3007M}). Unfortunately, little is known about the mass-ratio distribution of BNS systems due to insufficient data on the population of merging NS systems~\citep{2019ApJ...876...18F}. However, we can still obtain conservative estimates for the mass-radius posterior distribution for individual NSs considering a uniform distribution for the mass ratio and marginalizing over this variable when translating our posteriors on $\tilde{\Lambda}$ and $\mathcal{M}$ to posteriors on $M_{A}$ and $R_{A}$ for the $A^{\rm th}$ star in the binary. We show this result in Fig.~\ref{fig5}.

        We note that the results for the individual radii have larger uncertainties than our mass-radius constraint shown in Fig.~\ref{fig4} (as expected), since that constraint used information from both GRBs analyzed in this work. However, despite the lack of information on the binary mass ratio, our posteriors on the individual masses and radii have comparable uncertainties to the NICER pulsars (see, {\it e.g.},~\citealt{2019ApJ...887L..24M} and~\citealt{2021ApJ...918L..28M}) and GW170817~\citep{2018PhRvL.121p1101A}.

    \section{Discussion}
       
        We associate the frequencies of the QPOs in GRBs 910711 and 931101B reported by~\cite{2023Natur.613..253C} with BNS postmerger oscillation modes and obtain constraints on the redshift of these GRBs, as well as on the chirp mass and binary tidal deformability of the BNS systems whose mergers were presumably their source. We use the redshift to estimate the intrinsic oscillation frequencies of the fundamental quadrupolar mode and the quasiradial mode of the merger remnant in each case, and the chirp mass and tidal deformability to estimate the radius of a 1.4 M$_{\odot}$ NS and constrain the mass-radius relation. Therefore, our study introduces a novel way to constrain the EOS of NS matter, using QPOs from short GRBs. Using only two detections, we are able to estimate the radius of a $1.4$ M$_{\odot}$ neutron star as $R_{1.4}=12.48^{+0.41}_{-0.40}$ km.
        
        It is important to acknowledge as a caveat that the correspondence between frequencies of QPOs in short GRBs and those in the GW spectrum of BNS mergers is not fully certain. For instance, \cite{2023ApJ...947L..15M} show that strongly magnetized HMNSs can launch mildly relativistic outflows with kilohertz quasiperiodic features, although these features are not necessarily correlated with the postmerger GW signal. In a subsequent study, however, \cite{2024arXiv240401456M} show that the collapse of a rotating magnetar launches an outflow that is modulated by the ringdown of the black hole. Additional modelling and simulations are needed to clarify this scenario.

        The origin of the QPOs found in~\cite{2023Natur.613..253C} seems to be indeed short GRBs produced by BNS mergers, rather than, {\it e.g.}, giant flares from magnetars. If the GRB is launched by the HMNS, the flux can be modulated by the HMNS oscillations through the magnetic field~\citep{2019ApJ...884L..16C}. However, different sources for the frequencies of the QPOs are still possible, such as the turbulence on the accretion disk around the remnant. In such cases, however, it would be challenging to obtain coherent high-frequency emission due to destructive interference in the GRB light curve from turbulence fluctuations.
        
        Future detections of GWs with ground-based detectors such as LIGO, Virgo, and KAGRA, in coincidence with GRBs with QPOs, will allow a test of our model because the GW event will provide independent measurements of the redshift, chirp mass and binary tidal deformability, even if the postmerger signal is not  detectable. In the longer term, similar detections with next-generation ground-based detectors will allow for a direct comparison between the postmerger and QPO frequencies. Another way to test our model will be through the measurement of the time delay between the GW and GRB signals, which will allow us to put constraints on the time delay between the merger and the jet-launching, and the nature of the central engine (see, {\it e.g.}, \citealt{2019ApJ...876..139G, 2021ApJ...908..152M, Gutierrez:2024pch}). 

        Finally, we note that the QPOs in~\cite{2023Natur.613..253C} also have widths, which can be associated with the damping time of the quadrupolar and quasiradial modes. While the quadrupolar mode is also damped due to the emission of GWs, the quasiradial mode is solely damped through viscosity processes, which play an important role in the stability of remnant \citep{2018PhRvL.120d1101A}. Therefore, using the QPO widths as a proxy for or at least a constraint on the damping times of these modes could probe models for the effective viscosity in the remnant.

    \begin{acknowledgments}
        The authors thank Cole Miller and Jim Lattimer for useful discussions. V.G. acknowledges funding from the Southeastern Universities Research Association for the CRESST II/NASA Summer Internship at NASA/GSFC, during which part of this work was completed. D.R. acknowledges funding from the U.S. Department of Energy, Office of Science, Division of Nuclear Physics under Award Number(s) DE-SC0021177 and DE-SC0024388, and from the National Science Foundation (NSF) under Grants No. PHY-2011725, PHY-2020275, PHY-2116686, PHY-2407681, and AST-2108467. C.C. acknowledges support by NASA under award numbers 80GSFC21M0002 and 80NSSC20K0288. K.Y. acknowledges support from NSF Grant PHYS-2339969 and the Owens Family Foundation. 
    \end{acknowledgments}

    \paragraph{Note} During the review of this manuscript, the work from~\cite{2025ApJ...980..220H} was announced on arXiv. The authors provided two alternative interpretations for the QPOs reported in~\cite{2023Natur.613..253C}: torsional oscillations of magnetars and radial oscillations of hot neutron stars.

    \bibliography{references}

\end{document}